\documentclass[fleqn,10pt]{wlpeerj} 

\usepackage{empheq}
\usepackage[english]{babel}
\usepackage{amsfonts}
\usepackage{amsmath}
\usepackage{amssymb} 
\usepackage{graphicx} 
\usepackage[ruled,linesnumbered]{algorithm2e}
\usepackage{float}
\usepackage{placeins}
\usepackage{url, soul}
\usepackage{xcolor}
\usepackage[autolanguage]{numprint}
\usepackage{adjustbox}
\usepackage{tikz}
\usepackage{dsfont}
\usepackage{enumitem}
\usepackage{natbib}
\usepackage{tcolorbox}
\tcbuselibrary{breakable}
\graphicspath{{./figures/}} 

\colorlet{greendiff}{green!30}

\colorlet{reddiff}{red!30}

\colorlet{orangediff}{orange!20}

\colorlet{bluediff}{blue!30}

\title{Learning the optimal scale for GWAS through hierarchical SNP aggregation}

\author[1,2,*]{Florent Guinot}
\author[1]{Marie Szafranski}
\author[1,3]{Christophe Ambroise}
\author[1]{Franck Samson}
\affil[1]{UMR 8071 LaMME -- UEVE, CNRS, ENSIIE, USC INRA -- 91000 \'Evry, FRANCE}
\affil[2]{BIOptimize -- 51000 Reims, FRANCE}
\affil[2]{UMR MIA-Paris -- AgroParisTech, INRA, Universit\'e Paris-Saclay -- 75005 Paris, FRANCE}

\affil[*]{To whom correspondence should be adressed: florent.guinot@genopole.cnrs.fr}

\begin{abstract}
\textbf{Motivation:} Genome-Wide Association Studies (GWAS)
  seek to identify causal genomic variants associated with rare human
  diseases. The classical statistical approach for detecting these
  variants is based on univariate hypothesis testing, with healthy
  individuals being tested against affected individuals at each locus.
  Given that an individual's genotype is characterized by up to one
  million SNPs, this approach lacks precision, since it may yield a
  large number of false positives that can lead to erroneous
  conclusions about genetic associations with the disease. One way to
  improve the detection of true genetic associations is to reduce the
  number of hypotheses to be tested by grouping SNPs.

  ~\newline \textbf{Results:} We propose a dimension-reduction
  approach which can be applied in the context of GWAS by making use
  of the haplotype structure of the human genome. We compare our
  method with severals approaches on both synthetic and real GWAS
  data, and we show that reducing the dimension of the predictor
  matrix by aggregating SNPs gives a greater precision in the
  detection of associations between the phenotype and genomic
  regions. ~\newline
  
\end{abstract}

\begin{document}

\flushbottom
\maketitle
\thispagestyle{empty}

\section*{Introduction}

\subsection*{Context}

Recent breakthroughs in microarray technology have meant that hundreds
of thousands of single nucleotide polymorphisms (SNPs) can now be
densely genotyped at moderate cost. As a result it has become possible
to characterize the genome of an individual with up to a million
genetic markers. These rapid advances in DNA sequencing technologies
have also made it possible to carry out exome and whole-genome
sequencing studies of complex diseases. In this context, Genome-Wide
Association Studies (GWAS) have been widely used to identify causal
genomic variants implied in the expression of different human diseases
(rare, Mendelian, or multifactorial diseases).

From a statistical point of view, looking for these variants can be
supported by hypothesis testing. The standard approach in GWAS is
based on univariate regression (logistic regression in case-control
studies), with affected individuals being tested against healthy
individuals at one or more loci. Classical testing schemes are subject
to false positives (that is to say SNPs that are falsely identified as
significant variables). One way around this problem is to apply a
correction for the False Discovery Rate
\cite[FDR,][]{benjamini_controlling_1995,
  dalmasso_simple_2005}. Unfortunately, this increases the risk of
missing true associations that have only a small effect on the
phenotype, which is usually the case in GWAS.
\cite{maher_personal_2008} suggested that standard approaches such as
multiple hypothesis testing may not be appropriate for the detection
of small effects from multiple SNPs. In such cases a significant part
of the heritability can be missing and GWAS fails to detect all
possible genetic variants associated with a disease.

Furthermore, this kind of standard approach faces other limitations:
\vspace{-\topsep}
\begin{enumerate}[noitemsep]
\item It does not directly account for correlations among the
  predictors, whereas these correlations can be very strong as a
  result of linkage disequilibrium (LD).  SNPs can be correlated even
  where they are not physically linked, because of population
  structure or epistasis (gene by gene interactions).
\item It does not account for epistasis, i.e. causal effects that are
  only observed when certain combinations of mutations are present in
  the genome.
\item It does not directly provide predictive models for estimating the genetic
  risk of the disease.
\item \label{lim:last} It focuses on identifying common markers with
  allele-frequency (MAF) above 5$\%$, although it is likely that
  analyzing low-frequency ($0.5\% <$ MAF $< 5\%$) and rare (MAF $<
  0.5\%$) variants would be able to explain additional disease risks
  or trait variability~\citep{lee_rare-variant_2014}.
\end{enumerate}

Uncovering some of the missing heritability can sometimes be achieved
by taking into account correlations among variables, interaction with
the environment, and epistasis, but this is rarely feasible in the
context of GWAS because of the multiple testing burden and the high
computational cost of such analyses~\citep{manolio_finding_2009}.
In the context of rare-variant association analysis, a number
of region- or gene-based multimarker tests have been proposed as
burden tests~\citep{asimit_ariel_2012}, variance-component
  tests~\citep{wu_rare-variant_2011} or combined burden and variance
  component tests~\citep{lee_optimal_2012}. Instead of testing each
  variant individually, these methods evaluate the cumulative effects
  of multiple genetic variants in a gene or a region, increasing power
  when multiple variants in the group are associated with a given
  disease or trait.

Furthermore, regarding limitation~\eqref{lim:last},
  analyzing rare variants is more complex than analyzing more common
  variants and a large sample size is needed to observe a rare variant
  with a high probability.

\subsection*{Group and aggregation based methods for common
    variants}

Although classical GWAS have limitations that prevent a full
understanding of the heritability of genetic and/or multifactorial
diseases, there are nevertheless ways of overcoming these limitations
to some degree. For instance, it is possible to take into account the
structure of the data in the hypothesis testing procedure. As an
  illustration, 

\cite{meinshausen_hierarchical_2008} proposed a hierarchical testing
approach which considers the influence of clusters of highly
correlated variables rather than individual variables. The statistical
power of this method to detect relevant variables at single SNPs level
was comparable to that of the Bonferroni-Holm procedure
\citep{Holm_Bonferroni}, but the detection rate was much higher for
small clusters, and it increased further at coarser levels of
resolution.

Group-based methods require an appropriate group definition. In GWAS,
the usual approach is to group SNPs which are included in the same
gene but this limits the analysis to coding regions. It is well known
that the human genome is structured into haplotype blocks,
i.e. sizable regions over which there is little evidence for
historical recombination and within which only a few common haplotypes
may be observed~\citep{ardlie_patterns_2002}. The boundaries of blocks
and the specific haplotypes that they contain are highly correlated
across populations~\citep{gabriel_structure_2002}.  With this property
of the human genome in mind, \cite{huang_detecting_2007} developed a
method for detecting haplotype-disease associations in genome-wide
studies, based on sliding windows of adjacent SNPs, along with a Monte
Carlo procedure to adjust for multiple testing.

In \cite{wu_powerful_2010}, the authors proposed to group SNPs into
sets on the basis of their proximity to genomic features such as genes
or haplotype blocks and then to 
identify the joint effect of each set via a logistic
kernel-machine-based test.  This approach lays the foundation for the
Sequence Kernel Association Test method
\citep[SKAT,][]{wu_rare-variant_2011}.

In the broad family of linear models, \cite{listgarten_powerful_2013}
introduced a likelihood ratio-based set test that accounts for
confounding structure. The model is based on the linear mixed model
and uses two random effects, one to capture the set association signal
and one to capture confounders. They demonstrate a control of type I
error as well as an improved power over more traditionally used score
test.
Other methods focus on multiple linear regression either by taking
into account the linkage disequilibrium within the genes to improve
power \citep{yoo_multiple_2016} or by clustering variants with weak
association around known loci to increase the percentage of variance
explained in complex traits \citep{pare_contribution_2015}.

Finally, other approaches will focus on the aggregation of summary
statistics of single SNPs within a same gene with for instance the
data-driven aggregation of summary statistics described in
\cite{kwak_adaptive_2016} or the procedures of $p$-value combination
in \cite{petersen_assessing_2013}.
In the cited articles, the methods are used on SNPs located in coding region (or extended intronic region in~\cite{petersen_assessing_2013}) but can be extended to any set of SNPs as long as we pre-specified a set of variants within a region. However the power for each test remains dependent of the true disease model. 
Furthermore, this kind of approaches may also lose statistical power
in comparison to single-variant-based tests when only a very small
number of the variants in a gene are associated with the trait, or
when many variants have no effect or causal variants are low-frequency
variants~\citep{lee_rare-variant_2014}.

\subsection*{Organisation of the paper}
The present paper proposes a block-wise approach for GWAS analysis
which leverages the LD structure among the genomic variants to reduce
the number of hypotheses testing. We aggregate the SNPs into different
clusters according to their LD levels and use a supervised learning
approach to identify the clusters of SNPs related to a case-control
phenotype. Our algorithm provides a group structure for the variables,
enabling us to define a function that aggregates these clusters into
new variables to be used in the GWAS machinery. The advantage of this
method is that aggregating clusters of several SNPs into a single
variable reduces the dimension of the data without loss of
information, since we are grouping variables that are highly
correlated (in strong LD). 

We compare our method in different scenarios with the baseline
approach, i.e. univariate hypothesis testing
\citep{purcell_2007_plink} and with a state-of-the-art method, the
logistic kernel machine method developed by \cite{wu_powerful_2010} on
both synthetic and real datasets from the Wellcome Trust Case Control
Consortium~\citep{burton_genome-wide_2007} and on ankylosing
spondylitis data~\citep{IGAS_2013_identification}.

\section*{Method}

\begin{figure}[h]
  \centering
  \fbox{\includegraphics[width=.8\textwidth]{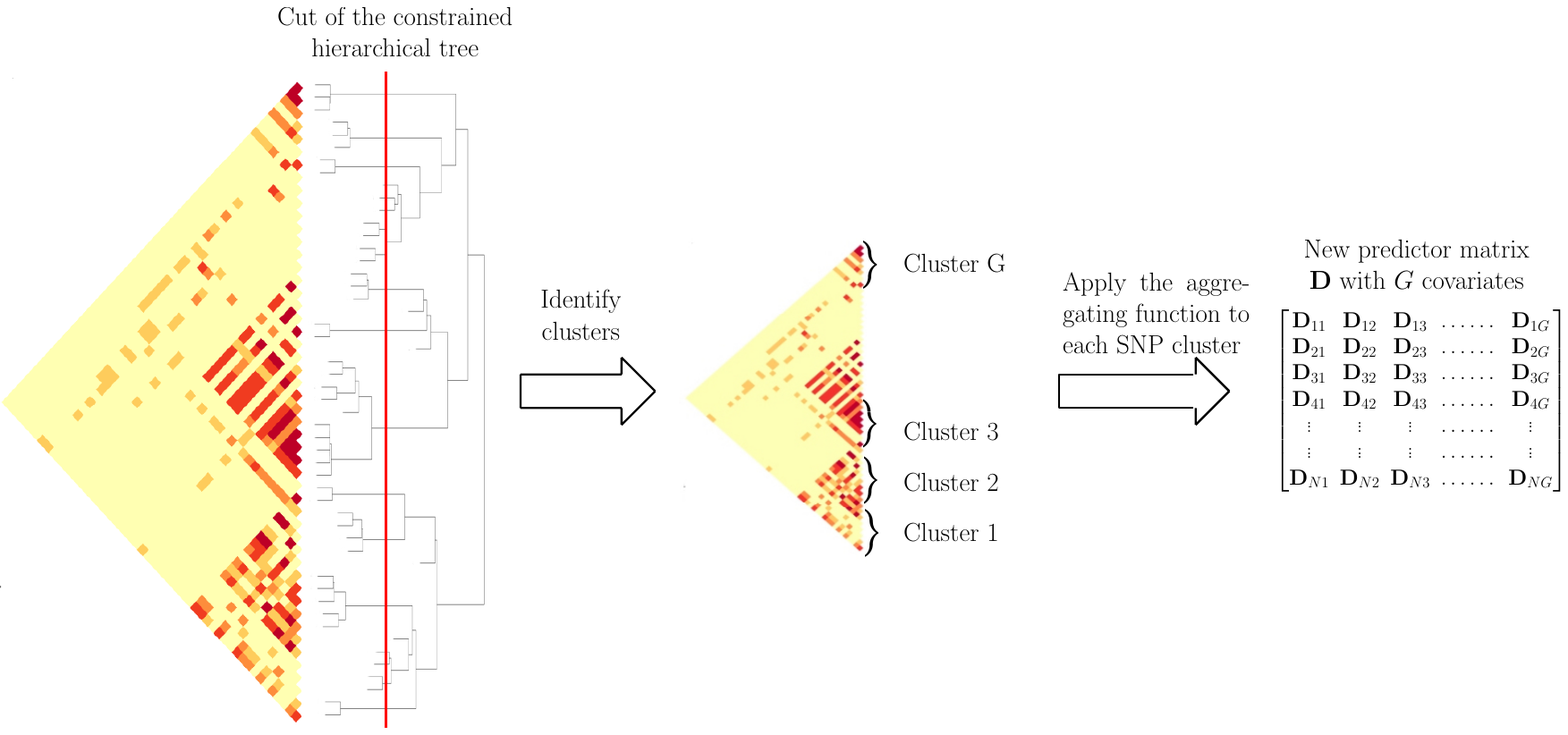}}
  \caption{Schematic view of Step 2 of the algorithm to calculate the
    matrix of predictors $\mathrm{D}$.}
  \label{fig:Step2}
\end{figure}

In this section we describe a new method for performing GWAS using a
four-step method that combines unsupervised and supervised learning
techniques. This method improves the detection power of genomic
regions implied in a disease while maintaining a good
interpretability. 
This method consists in:
\vspace{-\topsep}
\begin{enumerate}[noitemsep]
\item Performing a spatially constrained Hierarchical Agglomerative
  Clustering (constrained-HAC) of the SNPs matrix $\mathrm{X}$ using
  the algorithm developed by \cite{dehman_performance_2015}.
\item Applying a function to reduce the dimension of $\mathrm{X}$
  using the group definition from the constrained-HAC. This step is
  described and illustrated in Figure \ref{fig:Step2}.
\item Estimating the optimal number of groups using a supervised
  learning approach to find the best cut into the hierarchical tree
  (cut level algorithm). This algorithm combines Steps 1 and 2 into an
  iterative process.
\item Applying the function defined in Step 2 to each group identified
  in Step 3 to construct a new covariate matrix and perform multiple
  hypothesis testing on each new covariate to find significant
  associations with a disease phenotype $y$.
\end{enumerate}

\subsection*{Step 1. Constrained-HAC}
\label{ssec:CHAC}

In GWAS, the covariates are ordinal and correspond to SNP genotypes
such that $\mathrm{X}_{ij} \in \{0, 1, 2\}$ corresponds to the number
of minor alleles at locus $j \in \left[ 1,\dots, J \right]$ for
observation $i \in \left[ 1,\dots, N \right]$.

To take into account the structure of the genome in haplotype blocks,
we group the predictors (SNPs) according to their LD in order to
create a new predictor matrix which reflects the structure of the
genome. We first use the algorithm developed
by~\cite{dehman_performance_2015}, which clusters SNPs into adjacent
blocks. The clustering method is a spatially constrained hierarchical
clustering based on Ward's incremental sum-of-squares
algorithm~\citep{ward_hierarchical_1963}, in which the measure of
dissimilarity is not based on the Euclidean distance but rather on the
linkage disequilibrium between two SNPs: \(1 - r^2(j,j')\). The
algorithm also makes use of the fact that the LD matrix can be modeled
as block-diagonal by allowing only groups of variables that are
adjacent on the genome to be merged, which significantly reduces the
computation cost.
This algorithm is available via the \texttt{R} package called
\texttt{adjclust} on \url{https://cran.r-project.org/web/packages/adjclust}.

\subsection*{Step 2. Dimension reduction function}
\label{ssec:Dstar}

One way of addressing issues related to high-dimensional statistics
(and in particular the multiple testing burden that we mentioned
above) is to reduce the dimensionality of the predictor matrix
$ \mathrm{X} \in \mathbb{R}^{N \times J} $ by creating a reduced
matrix $ \mathrm{D} $ with new covariates that nevertheless remain
representative of the initial matrix. This means reducing the number
of predictors $J$ to $G \ll J$, with row $\mathrm{D}_{i.}$ the
$G$-dimensional vector of new predictors for observation $i$. In this
study we use a blockwise approach to construct a matrix of new
uncorrelated predictors $ \mathrm{D} \in \mathbb{R}^{N \times G} $,
with $G$ the number of groups in linkage disequilibrium identified via
the constrained agglomerative hierarchical clustering described in
Step 1.

While classical methods use the initial set of covariates to predict a
phenotype, we propose combining a clustering model with a dimension
reduction approach in order to predict $y$. For each group identified
with the constrained-HAC, we apply a function to obtain a single
variable defined as the number of minor alleles present in the group.
For each observation $i$ and in each cluster
$ g \in \left[1,\dots,G\right]$, the variable is defined as:
 \begin{equation}
  \label{eq:1}
  \mathrm{D}_{ig}~=~\sum_{j \in g} \mathrm{X}_{ij}.
\end{equation} 

In order that the values for the different groups are comparable, we
eliminate the effect of group size by centering and scaling the matrix
$\mathrm{D}$ to unit variance. In the remainder of the paper we will
refer to the covariates in $\mathrm{D}$ as \emph{aggregated-SNP}
variables.

\subsection*{Step 3. Optimal number of groups estimation}
\label{ssec:cutree}
\begin{algorithm}[h]
  \SetKwFor{For}{for}{do}{end}
  \SetKwData{hierarchy}{hierarchy}
  \SetKwData{cutlevel}{cutlevel}
  \SetKwData{ridgecoef}{ridgecoef}
  \SetKwData{AUC}{AUC}
  \SetKwData{bestlevel}{bestlevel}
  \SetKwData{Dtrain}{$D^{train}_{sub1}$}
  \SetKwData{Dtest}{$D^{test}_{sub1}$}
  \SetKwData{Ypred}{$Y^{pred}_{sub1}$}
  \SetKwData{Dbest}{$D^{best}_{sub1}$}
  \SetKwFunction{Max}{Max}
  \SetKwFunction{Aggregating}{Aggregating}
  \SetKwFunction{RidgeRegression}{RidgeRegression}
  \SetKwFunction{Predict}{Predict}
  \SetKwFunction{ROC}{ROC}
  \SetKwFunction{Which}{Which}
  \SetKwFunction{Sequence}{Sequence}
  \SetKwInOut{Input}{input}
  \SetKwInOut{Output}{output}
     
  \Input{Covariates matrix $\mathrm{X}_{sub1}$} \Output{Matrix
    $\mathrm{D}^{best}_{sub2}$ of aggregated-SNPs at best cut level}
  \BlankLine
  \emph{Define training and test set}\;
  \hierarchy$\leftarrow$ Constrained-HAC on $\mathrm{X}^{train}_{sub1}$\\
  \cutlevel$\leftarrow$ Initialize levels where to cut hierarchy\\
 
  \For{$i\leftarrow$ \Sequence{\cutlevel}}{
    \Dtrain$\leftarrow$ \Aggregating{$\mathrm{X}^{train}_{sub1}$,
      \hierarchy, $\cutlevel[i]$}\;
    \Dtest $\leftarrow$ \Aggregating{$\mathrm{X}^{test}_{sub1}$,
      \hierarchy, $\cutlevel[i]$}\;
    \ridgecoef$\leftarrow$ \RidgeRegression{$\mathrm{Y}^{train}_{sub1}
      \thicksim 
      \mathrm{D}^{train}_{sub1}$}\;
    \Ypred$\leftarrow$ \Predict{$\mathrm{X}^{test}_{sub1}$,
      \ridgecoef}\;
    $\AUC[i]\leftarrow$ \ROC{$\mathrm{Y}^{test}_{sub1},
      \mathrm{Y}^{pred}_{sub1}$}\;
  }
  \bestlevel$\leftarrow$ \Which{\cutlevel, \Max{\AUC}} \;
  \Dbest$\leftarrow$ \Aggregating{$\mathrm{X}_{sub1}$, \hierarchy,
    \bestlevel}\; 
  \caption{Supervised learning cut level algorithm}
  \label{algo_cutlevel}
\end{algorithm}

Estimating the optimal number of groups to select, i.e. the level at
which the hierarchical clustering tree should be cut, is a fundamental
matter which impacts the relevance of the association analysis. It is
known that the human genome is structured into haplotype blocks with
little or no within-block recombination
\citep{gabriel_structure_2002}, but it is not easy to determine how
these blocks are allocated throughout the genome for a given set of
SNPs.

In the literature, in an unsupervised learning context, a number of
models have been proposed for determining this optimal number of
groups~\citep{tibshirani_estimating_2001, hartigan_clustering_1975,
  calinski_dendrite_1974, krzanowski_criterion_1988}. These methods
are all based on the measure of within-group dispersion $W_{G}$ with
$ G \in \left[1,\dots,P \right] $. Since GWAS consist in evaluating
the likelihood of the disease from genetic markers, we propose using
the phenotype $y$ as a way of determining the optimal number of
clusters.

We propose here a supervised validation set approach to find this
optimum.  Since this algorithm aims to identify phenotype-related SNPs
clusters, it is necessary to split the dataset into two subsets to
avoid an inflation of type I errors in the testing procedure. One
subset, $[Y_{sub1}, \mathrm{X}_{sub1}]$, is used to choose the optimal
cut and the second one, $[Y_{sub2}, \mathrm{X}_{sub2}]$, to perform
the hypothesis testing in Step 4.

First we apply the constrained-HAC on a subset
$\mathrm{X}^{train}_{sub1} \subset \mathrm{X}_{sub1}$, and for a given
level of the hierarchy we apply the dimension reduction function
defined above (Step 2) to each of the $G$ clusters to construct the
matrix $\mathrm{D}^{train}_{sub1}$. We then fit a ridge regression
model to estimate the coefficients of the predictors in
$\mathrm{D}^{train}_{sub1}$. Ridge regression is a penalized model
which shrinks the estimated coefficients towards zero and is known to
have a good stability in comparison to other penalized-regression
models such as lasso regression \citep{bousquet_stability_2002}.
Moreover, a link can be established between the ridge regression model
and the mixed linear model used in the estimation of the heritability
in a high-dimensional setting \citep{bonnet_heritability_2014}.  Once
the coefficients are estimated, we predict the phenotypic values on
the test set using the matrix $\mathrm{D}^{test}_{sub1}$ and calculate
either the mean test set error when the phenotype is quantitative or
the Area Under the ROC curve (AUC-ROC) when it is binary. The
procedure, summarized in Algorithm \ref{algo_cutlevel}, is then
repeated for different levels in the hierarchy and the optimal cut
level in the tree is defined as the level which maximizes the
prediction accuracy criterion.

\subsection*{Step 4. Multiple testing on aggregated-SNP variables}

Once the optimal number of groups has been determined, we apply the
function \eqref{eq:1} to each selected group and construct the matrix
of \emph{aggregated-SNP}. Here we use a standard Single Marker
Analysis (SMA) to find associations with the phenotype, but instead of
calculating $p$-value for each SNPs in $\mathrm{X}_{sub2}$, we
calculate $p$-value for each \emph{aggregated-SNP} variables in
$\mathrm{D}_{sub2}$.

As in standard SMA, a univariate generalized linear
model~\citep{nelder_generalized_1972} is fitted for each variable
$\mathrm{D}_{.j}$:\, $\mathit{f(\mu_i)} = \mathrm{D}_{ij}\beta, $
where $\mu_i \equiv \mathbb{E}(\mathrm{Y}_i|\mathrm{D}_i) $
($\mathrm{Y_i} \thicksim \text{some exponential family
  distribution})$, $f$ is a smooth monotonic 'link function',
$\mathrm{D}_{ij}$ is the $\textit{i}^{th}$ row of the model matrix
$\mathrm{D}_{.j}$ of aggregate-SNP and $\beta$ is a vector of 2
unknown coefficients with $\beta_0$ for the intercept and $\beta_1$
for the predictor $j$. Where the response variable is a binary trait
(i.e. case-control phenotype), we use the logit function as the 'link
function' $f$ and $ Y_i \thicksim \text{Bernoulli distribution}$. This
model is known as the logistic regression model. Then, for each
single-predictor model, we perform a Likelihood Ratio Test where we
compare the intercept-only model against the single-predictor model
and get for each predictor a $p$-value using the $\tilde\chi^2$
distribution.

Given that a large number of covariates are being tested, we need to
compute an appropriate significance threshold to control the
family-wise error rate, $FWER = \mathbb{P}(FP > 1)$, with $FP$ being
the False Positive, since keeping the threshold at the conventional
value of $\alpha = 0.05 $ would yield numerous false positives.
Several approaches, including the Bonferroni correction, have been
proposed in the context of genetic studies for controlling the FWER
~\citep{Sham_statistical_2014}. An alternative approach, developed by
\cite{benjamini_controlling_1995}, seeks to control the False
Discovery Rate (FDR) which is the expectation of ratio between the
number of false positives and the total positive outcomes:
$\displaystyle{FDR = \mathbb{E}\left(\frac{FP}{FP + TP}\right)}$, with
$TP$ being the True Positive.
The Bonferroni correction reduces the significance level according to
the number of tests carried out in the study. However, in the context
of GWAS, where hundreds of thousands of tests have to be performed,
the Bonferroni correction is too strong, and will often decrease the
significance threshold to a level where almost nothing is significant.
Controlling FDR is therefore preferable. It is an approach that is
less stringent but nonetheless powerful. The method for controlling
FDR does not directly set a significance threshold, but rather
identifies the largest $p$-value that is substantially smaller
than its expected value (by a factor of at least 1/$\phi$ where $\phi$
is the desired FDR level), given that all the tests follow $H_0$. The
$p$-value thus identified and all smaller $p$-values are
deemed to be significant.

\section*{Numerical simulations}
\label{sec:numsim}

The performance evaluation described below was designed to assess the
ability of our method to retrieve causal SNPs or causal clusters of
SNPs under different simulation scenarios.
For each scenario, we use a matrix $\mathrm{X_{HAPGEN}}$ of SNPs
generated by the \textsf{HAPGEN2} software \citep{su_hapgen2_2011}
with a sample size of 1000 individuals. This software allows to
simulate an entire chromosome conditionally on a reference set of
population haplotypes (from HapMap3) and an estimate of the fine-scale
recombination rate across the region, so that the simulated data share
similar patterns with the reference data. We generate the chromosome 1
(103 457 SNPs) using the haplotype structure of CEU population (Utah
residents with Northern and Western European ancestry from the CEPH
collection) as reference set. The HAPGEN2 software allows to generate
a controls-only matrix of SNPs (no disease allele).  We filtered this
matrix according to the minor allele frequency to only keep SNPs with
a MAF greater than $5\%$ thus reducing the size of
$\mathrm{X_{HAPGEN}}$ to 60 179 SNPs.

 We generate a posteriori the phenotype using the logit model with a given set of
causal SNPs or cluster of SNPs. The main difference between the
different scenarios is to be found in the way that the case-control
phenotype y is simulated.

\subsection*{Simulation of the case-control phenotype}
\label{ssec:simupheno}

For all simulation scenarios, we simulated a case-control phenotype
$y$ under a logistic regression model.  We chose a matrix
$\tilde{\mathrm{X}} $ with $\ell$ causal variables
to create the case-control phenotype $y$ under the logit model:
$$\mathbb{P}(y_i=1|\tilde{\mathrm{X}}_{i.}) = \frac{\exp(\beta_0 +
    \boldsymbol\beta \tilde{\mathrm{X}}_{i.})} {1 + \exp(\beta_0 +
    \boldsymbol\beta \tilde{\mathrm{X}}_{i.})}\;, $$
where $\boldsymbol\beta = [\beta_1, \dots, \beta_\ell] $ is the vector
of coefficients corresponding to the $\ell$ predictors
$[\tilde{\mathrm{X}}_{.1}, \dots, \tilde{\mathrm{X}}_{.\ell}]$ and
$\beta_0$ is the intercept defined as
${ ln\left(\frac{\pi}{(1-\pi)}\right)}$, with $\pi$ the true
prevalence of the disease in the population. The predictors are
centered to have zero-mean before generating the vector of
probability.

The logit function gives a vector of probabilities for the phenotype
equals to 1, conditionally on the $\ell$ predictors. The case-control
phenotype is then generated following a Bernoulli distribution
function with a probability equal to
$\mathbb{P}(y_i=1|\tilde{\mathrm{X}}_{i.})$.

As we cannot control the residual $\epsilon_i$ in the logit model to
generate the phenotype, one way to have an association between the
response and the predictors strong enough to be detected is to set
large $\boldsymbol\beta$ coefficients on the predictors. Indeed there
is a direct relationship between the odd ratio ($OR$) of a covariate
$\tilde{\mathrm{X}}_{i.}$ and its corresponding coefficients
$\boldsymbol\beta_i$ in the logistic regression model
\citep{Diaz_simple_2012} given by $OR_i = e^{(\boldsymbol\beta_i)}$.
In our simulations, the difficulty of the problem is linked to the
number of causal predictors used to generate $y$ and the $OR$ set to
each predictors.

To simulate different scenarios we considered the following
parameters:
\vspace{-\topsep}
\begin{enumerate}[noitemsep]
   \item Nature of the causal predictors: 
     \vspace{-\topsep}
     \begin{itemize}[noitemsep]
     \item \textbf{Clusters of SNP:} For each replicates,
       $\ell = 1, 2, 3$ genomic regions have been identified to be
       causal. These regions have been chosen among the matrix
       $\mathrm{X_{HAPGEN}}$ to have different levels of LD among the
       SNPs that compose them. The average correlation coefficient
       among the SNPs in these regions varies from $r^2=0.6$ to
       $r^2=0.85$ and the size of the region varies from 20 SNPs to 60
       SNPs. Once identified, the causal regions were aggregated using
       the function described in Step \ref{ssec:Dstar} to construct a
       matrix $\tilde{\mathrm{X}}$ of \emph{aggregated-SNPs} predictors used
       to generate the case-control phenotype. We will refer to
       this scenario as the \emph{SNPclus} scenario.

     \item \textbf{Single SNPs.} In this scenario the phenotype was
       simulated by directly using sampled SNPs from the same causal
       regions identified in the \emph{SNPclus} scenario. For each
       replicates, we chose 10
       individuals SNPs among each of these regions to construct a matrix
      $\tilde{\mathrm{X}}$ with $\ell = 10 ,20 ,30$ single SNPs
      predictors, depending on the number of causal regions . This matrix is
      then used to generate the case-control phenotype. The chosen
      SNPs have a MAF varying from $10 \%$ to $30\%$.  We will refer to this
       scenario as the \emph{singleSNP} scenario.
     \end{itemize}
   
   \item Number of causal predictors $\ell$ and number of replicates:

     We performed $5$ replicates for each combination $\ell \times$
     number of scenarios and we evaluate the average performance over
     these $5$ replicates. For each scenario we considered from 1 to 3
     causal genomic regions, thus, for \emph{SNPclus} scenario, we
     used up to 3 causal predictors, and for the \emph{singleSNP}
     scenario, up to $10 \times 3 = 30$ causal
     predictors to generate the case-control phenotype.
   
   \item Odd ratio ($\beta$ coefficients) of the causal predictors: 

     For the \emph{SNPclus} scenario we chose an equal OR of 2.7 for
     each causal aggregated predictors, corresponding to a
     $\boldsymbol\beta$ coefficient equal to 1.  For the
     \emph{singleSNP} scenario we chose an equal OR of 1.1 for each
     causal predictors, corresponding to a $\boldsymbol\beta$
     coefficient equal to $0.1$. The rationale behind these
     coefficients arises from the hypothesis that the combined effect
     of several low-effect SNPs on the phenotype is stronger than the
     effects of each individual SNP.
\end{enumerate}

As previously mentionned, we generated the phenotype using causal SNPs
simulated with the HAPGEN2 software.  However, as commercial genechips
such as Affymetrix and Illumina arrays do not genotype the full
sequence of the genome, some SNPs are thereby unmapped and the marker
density is in general lower than the HapMap marker density. That is
why we chose, in our numerical simulation, to generate the phenotype
with causal variables chosen from the HAPGEN matrix and to assess the
performance of the methods using only those SNPs which are mapped on a
standard Affymetrix genechip (about 23 823 mapped SNPs). By doing so,
some causal SNPs are not mapped on the commercial SNP set and thus
simulations are more similar to real genome-wide analysis conditions.

\subsection*{Performance evaluation}
\label{ssec:perfeval}

\paragraph{Competitors.} 
The objective of our method being to identify the optimal scale at
which to perform association studies, we compared our proposal  with
several methods working at different genomic scales. The purpose is to
assess the ability of each method to retrieve true causal genomic
regions in the different simulation scenarios. For each scenario,  
four approaches have been considered: 
\begin{itemize}
\item SKAT\textit{tree}, a SKAT model using our group definition, 
\item SKAT\textit{notree}, a SKAT model using an alternative group definition produced by successive chunks of 20 SNPs,
\item SMA, the classical Single Marker Analysis,
\item SASA (Single \emph{Aggregated-SNP} Analysis) a method close to
  SMA, where instead of testing the genotype-phenotype association
  using each single SNP, we are testing it using \emph{aggregated-SNP}
  variables.
\end{itemize}
The two above described  group definitions for SKAT were considered 
to evaluate the impact of the group structure on the association findings.

The comparison with SMA allows to  highlight the advantage of working at 
a group scale. We hypothesize that  grouping low-effect SNPs should have a
better statistical power than  testing the main effects at single-SNP level.

For all methods, we compare the results using 2 types of multiple
testing corrections : the methods of Holm-Bonferroni
\citep{Holm_Bonferroni} and \cite{benjamini_controlling_1995}.

\paragraph{True and False Positive definitions.}
The problem of retrieving true causal associations can be represented
as a binary decision problem where the compared methods are considered
as classifiers. The decision made by a binary classifier can be
summarized using four numbers: True Positives ($TP$), False Positive
($FP$), True Negatives ($TN$) and False Negatives ($FN$).  We
represent True Positive Rate ($\text{Recall or Power} ={TP/(FN+TP)}$)
versus Precision ($\text{Precision}={TP/(FP+TP)}$).  

In this context, a True Positive corresponds to a true
causal genomic region associated to significant \textit{p}-value.
The definition of what can be considered as the true causal genomic
region may nevertheless be subject to some ambiguity. In GWAS, the
presence of LD between SNPs often leads to consider the signal
associated to multiple neighboring SNPs as indicating the existence of
a single genomic locus with possible influence on the phenotype.

In our simulations, a causal genomic region is defined \emph{a priori}
as a causal predictor in the logit model. However, since the clusters
of SNPs identified by our algorithm are not totally independent, some
residual correlation may remain between clusters. This leads to
question the notion of relevant variable when the variables are
structured into strongly correlated groups.  Should all the variables
of the block be considered as explanatory, or should we define as only
true positives the causal variables used to generate the phenotype ?

In order to circumvent this issue, we chose to relax the definition of
a False Positive joining the work of \cite{brzyski_controlling_2017}
and \cite{yi2015penalized} where they propose to control the FDR in
GWAS by considering significiant SNPs correlated to the true causal
variables as true positives.

For the simulation of the phenotype, we hypothesize an underlying
multivariate regression model, but test for univariate model as it is
the usual practice, which
leads to reconsider the definition of true positive. As in
\cite{yi2015penalized} we consider the set of true positive as the
union of the causal true positive and the linked
true positive, which are regions adjacent to the causal regions and
correlated with them at a level of at least 0.5.

Regarding the single-marker analysis approach, since it works at the
single SNP level, we compare it with the others in the
\emph{singleSNP} scenario only. 

\subsection*{Results and discussions of the numerical simulations} 
\paragraph{Area Under the ROC Curve.}

\begin{figure}[H]
  \centering
  \fbox{\includegraphics[width=.85\textwidth]{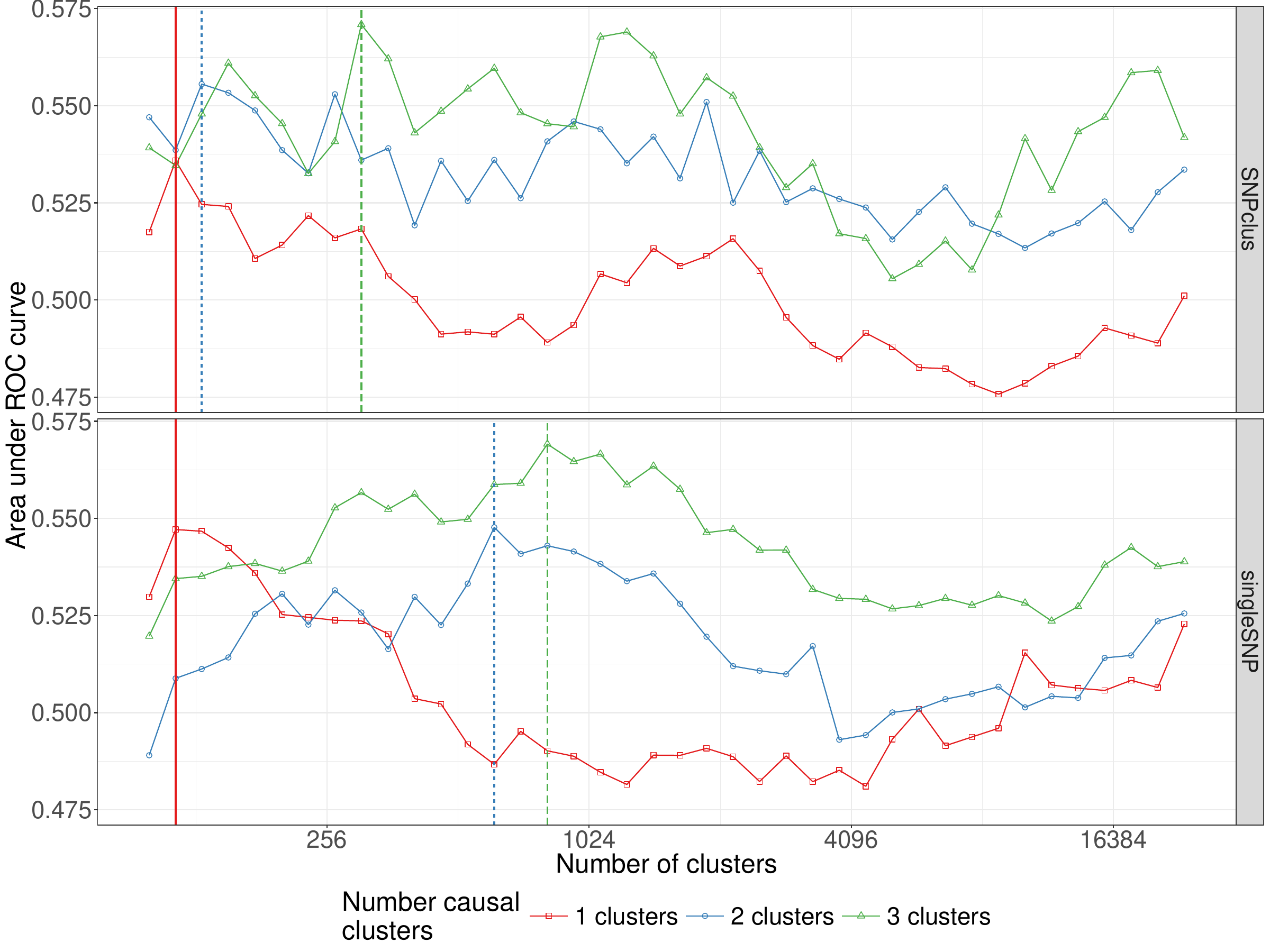}}
  \caption{Area under ROC curves according to the number of clusters in the
    \emph{clusSNP} and \emph{singleSNP} scenarios: the vertical lines indicate the number
    of aggregated-SNPs (clusters) obtained
    with~Algorithm~\ref{algo_cutlevel}, i.e. the level where the
    prediction error is minimized (AUC-ROC at its maximum).}
  \label{fig:AUC_curve}
\end{figure}

For each simulation, the cut level algorithm was applied. We recall
that this algorithm calculates a prediction error on a test set for
several levels in a constrained-HAC tree with a ridge regression model
and chooses the level for which this error is the smallest. The
AUC-ROC is plotted for the different levels, and the best cut level
corresponds to the level for which AUC-ROC is the greatest.  The
results from the simulation scenario \emph{clusSNP} and
\emph{singleSNP} described in Section \ref{ssec:simupheno} are shown
in Figure \ref{fig:AUC_curve}.
Our algorithm cuts the hierarchy either at a fairly high level (few
large clusters) or at a low level (many small clusters), depending on
the number of causal variables we used to generate the
phenotype. The more the number of causal regions decreases,
  the higher the algorithm cuts in the hierarchical tree.
In either case our algorithm is able to increase the predictive power
by aggregating SNPs with the function (1).

We are thus able build a matrix of uncorrelated aggregated-SNP
predictors that are representative of the initial SNP matrix and
strongly linked to the phenotype.

\paragraph{Performance results for simulated data. }
As described in Section \ref{ssec:perfeval}, we evaluate and compare
the methods using two metrics, namely \textit{Recall} and
\textit{Precision}. 

Here the \textit{Precision} metric is somewhat relaxed compared to its
true definition since we adapted the definition of a True Positive and
False Positive to the GWAS context. It is important to note that for
all the methods, we compare the Benjamini-Hochberg method to control
FDR with the Bonferroni correction to control FWER at a threshold of
$5\%$. However, since there are residual correlations between SNPs
clusters and that the replication of numerous samples per combination
of parameters is difficult in this realistic setting of simulations,
the observed Type I error rate may be greater than $5\%$.  What we
think is important to put forward to in these simulations is the
ability of our algorithm to define groups of relevant clusters that
will be detected on average with more precision and more power (SASA
and SKAT\emph{tree}) than using an arbitrary group definition
(SKAT\emph{notree}) or no definition of groups at all (SMA).

\begin{figure}[H]
  \centering
  \fbox{\includegraphics[width=.8\textwidth]{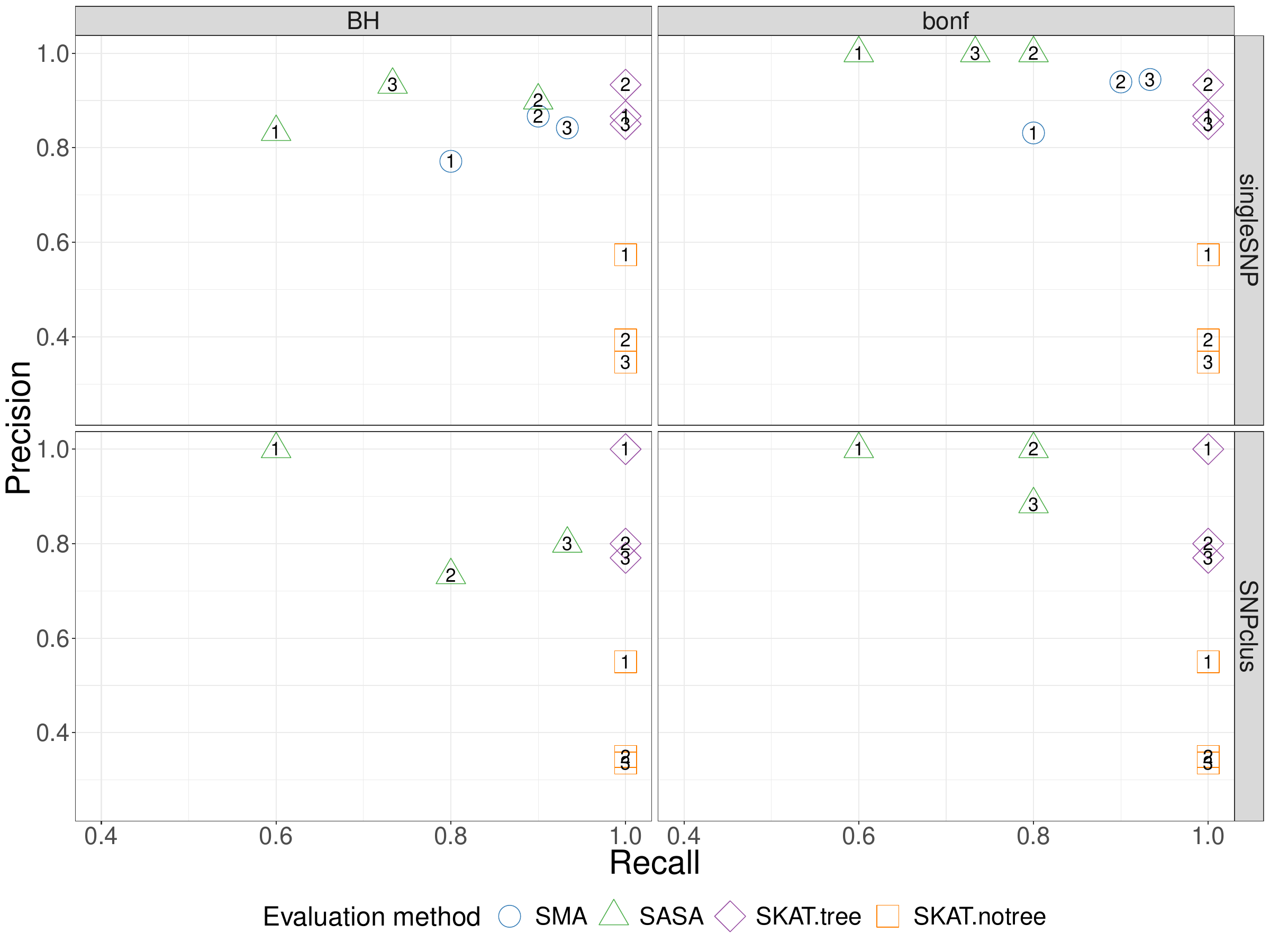}}
  \caption{Recall vs Precision for each method (shape and
      colors in plot). In rows are the simulation scenarios. In
      columns, we evaluate performance using Benjamini-Hochberg
      threshold (left) and bonferroni correction threshold (right).
      The second row illustrates the performance to retrieve the true
      causal genomic region under the \emph{SNPclus} scenario, thus
      only group-based approaches are considered (SASA, SKAT.tree and
      SKAT.notree). The numbers inside the points correspond to the
      number of causal predictors and each point is the average value
      of 5 replicates.}
  \label{fig:scatter_perf}
\end{figure}

The results represented in Figure \ref{fig:scatter_perf} show that the
methods using our algorithm for the cluster definition (SASA and
SKAT\textit{tree}) have in average a better precision than the two
other methods. The approach SASA, which combine our clustering
algorithm and the aggregating function (1) to test the association of
aggregated-SNPs with the phenotype, perform poorly in term of Recall
but is far better in term of Precision compared to SMA and
SKAT\textit{notree}. These results suggest that it is better to
combine our algorithm with the SKAT method than with the SASA method.
We also note that applying the SKAT approach on an arbitrary group
definition (SKAT\textit{notree}) lead to a good recall but a very poor
precision, showing the benefit of using our custom group definition in
this context. Regarding the SMA approach in the \emph{singleSNP}
scenario, we can observe a loss in term of Recall compare to the
SKAT\textit{tree} and SKAT\textit{notree} method suggesting that we
can take benefit of grouping low effect SNPs to improve the power to
detect causal genomic regions.

In GWAS, having a method with a good precision is as important, or
even more important, than having a good recall. It is better to spot a
few significant associations with a high certainty than to spot
numerous significant associations but with only a low level of
certainty for most of them. For this reason, we believe that our
method represents an improvement in terms of precision without loss of
power insofar as SKAT\textit{tree} seems able to detect significant
genomic regions associated with the phenotype with a higher degree of
certainty than standard approaches.

\section*{Application on real datasets}

To evaluate the performance of our method on real data, we performed
GWAS analysis on datasets made available
by~\citep{burton_genome-wide_2007}. The WTCCC data collection contains
17000 genotypes, composed of 3000 shared controls and 14000 cases
representing 7 common diseases of major public health concern:
inflammatory bowel disease (IBD), bipolar disorder (BD), coronary
artery disease (CAD), hypertension (HT), rheumatoid arthritis (RA),
and Type I (T1D) and Type II (T2D) diabetes.  Individuals were
genotyped with the Affymetrix GeneChip 500K Mapping Array Set and are
represented by about 500,000 SNPs (before the application of quality
control filters). 
In parallel to the analysis of the WTTCC data, we decided to assess
our method on another dataset from a different study. The ankylosing
spondylitis (AS) dataset consists of the French subset of the large
study of the International Genetics of Ankylosing Spondylitis (IGAS)
study \citep{IGAS_2013_identification}. For this subset, unrelated
cases were recruited through the Rheumatology clinic of Ambroise Paré
Hospital (Boulogne-Billancourt, France) or through the national
self-help patients’ association: "Association Française des
Spondylarthritiques". Population-matched unrelated controls were
obtained from the "Centre d’Etude du Polymorphisme Humain", or were
recruited as healthy spouses of cases. The dataset contains 408 cases
and 358 controls, and each individual was genotyped for 116,513 SNPs
with Immunochip technology.

\begin{figure}[H]
  \centering
  \fbox{\includegraphics[width=.7\textwidth]{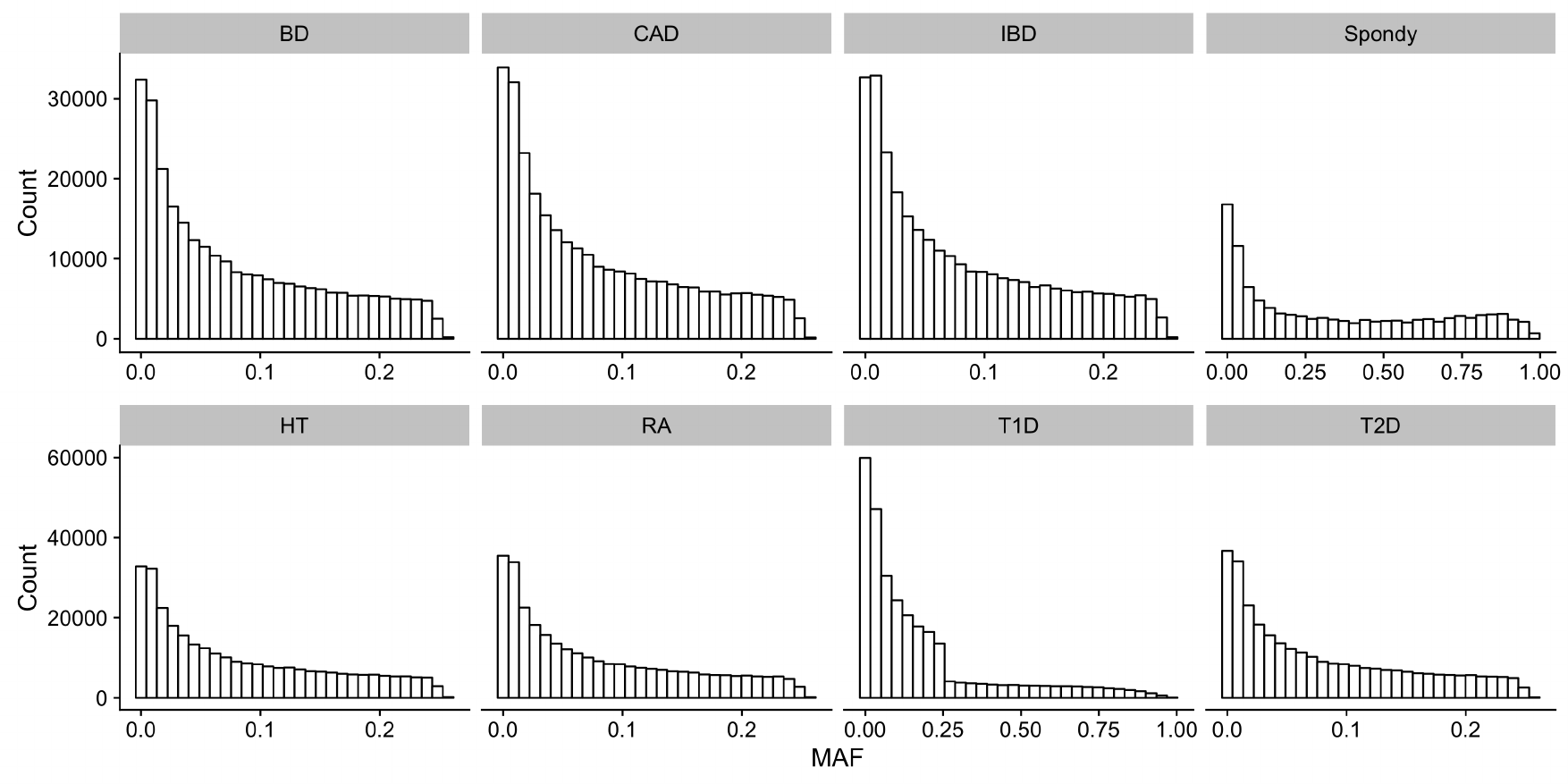}}
  \caption{Histograms of Minor Allele Frequencies
      (MAF) distribution in each datasets. (BD) Bipolar disorders;
      (CAD) Coronary arthery disease ; (IBD) Inflammatory bowel
      disease ; (HT) Hypertension ; (RA) Rheumatoid arthritis ; (T1D)
      Type I diabetes ; (T2D) Type II diabetes.}
  \label{fig:histomaf}
\end{figure}

To remove the bias induced by population stratification in Genome-Wide
analysis, we added the first 5 genomic principal components into the
regression model as described in \citep{price_principal_2006}. Since
the methods evaluated here do not deal with missing values, we chose
to impute the missing genotypes with the most frequent genotypic
value, $h_j$ observed for each $j$ SNP.  For each each dataset, we filtered the values to keep only those SNPs having a MAF greater than $5\%$. The minor allele frequencies of each datasets are represented in Figure \ref{fig:histomaf}.

We applied our cut level algorithm to find relevant clusters of SNPs
and we performed single marker analysis on single SNPs (SMA) and on
groups of SNPs (SASA, SKAT\textit{tree}, SKAT\textit{notree}).  We
then compared the significant associations detected by the different
methods to reveal possible new associations with the phenotype.

\subsection*{Results on real datasets}
\subsubsection*{AUC-ROC curves}

\begin{figure}[H]
  \centering
  \fbox{\includegraphics[width=.85\textwidth]{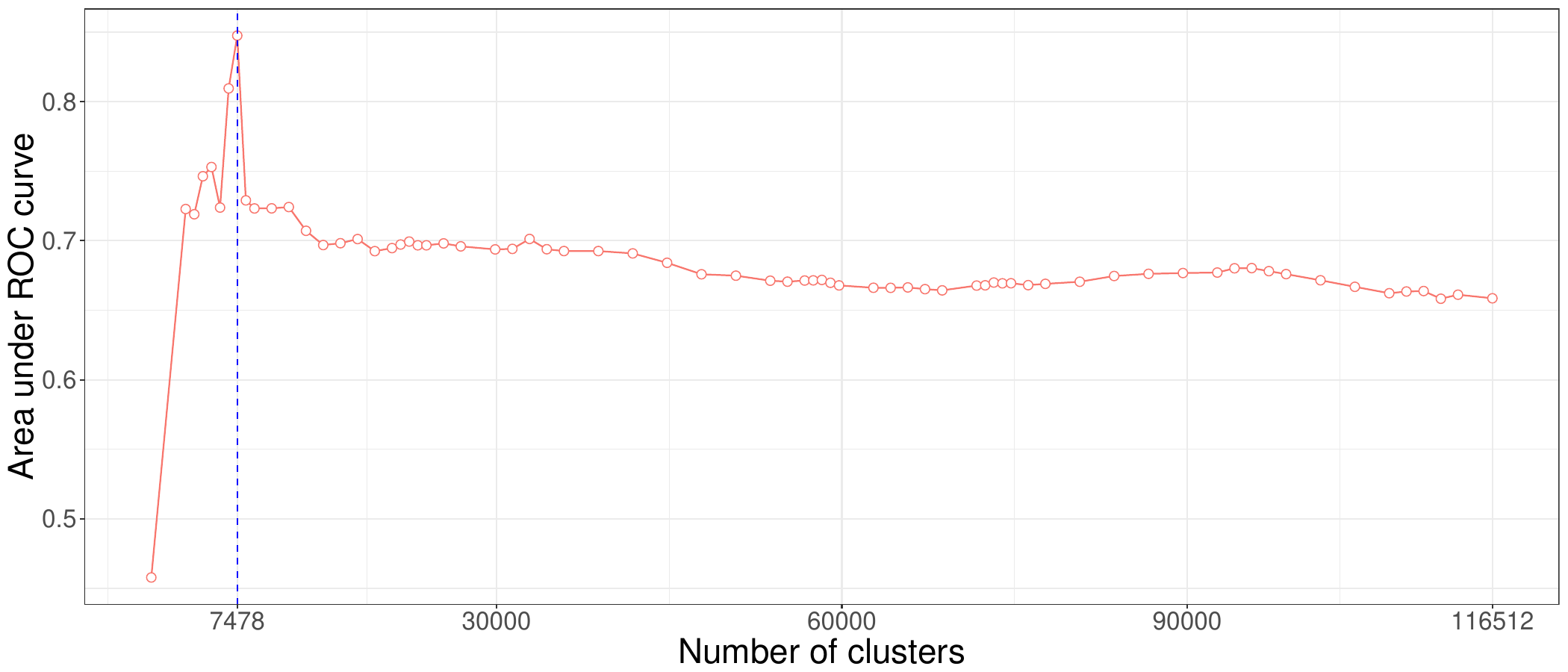}}
  \caption{AUC-ROC for different cut levels in a HAC-tree of 7
    WTCCC diseases after quality control filters. Each point
    corresponds to an AUC value computed on a test set from a
    logistic ridge regression model for a given level in the
    constrained-HAC tree.}
  \label{fig:WTCCC}
\end{figure}

In this section, we compare the AUC-ROC curves generated by our
cut level algorithm for each disease (WTCCC and AS data). Concerning the WTCCC diseases, given that patients were all genotyped
using the same GeneChip, their genotypes have the same LD structure,
and therefore the shapes of the AUC-ROC curves should be very similar
between the different diseases. As can be observed in
Figure~\ref{fig:WTCCC} (WTCCC diseases), the shape of the AUC-ROC
curves are closely similar, with a chosen cut level located around 100
000 clusters of SNPs, suggesting a shared LD pattern among patients.

\begin{figure}[H]
  \centering
    \fbox{\includegraphics[width=.8\textwidth]{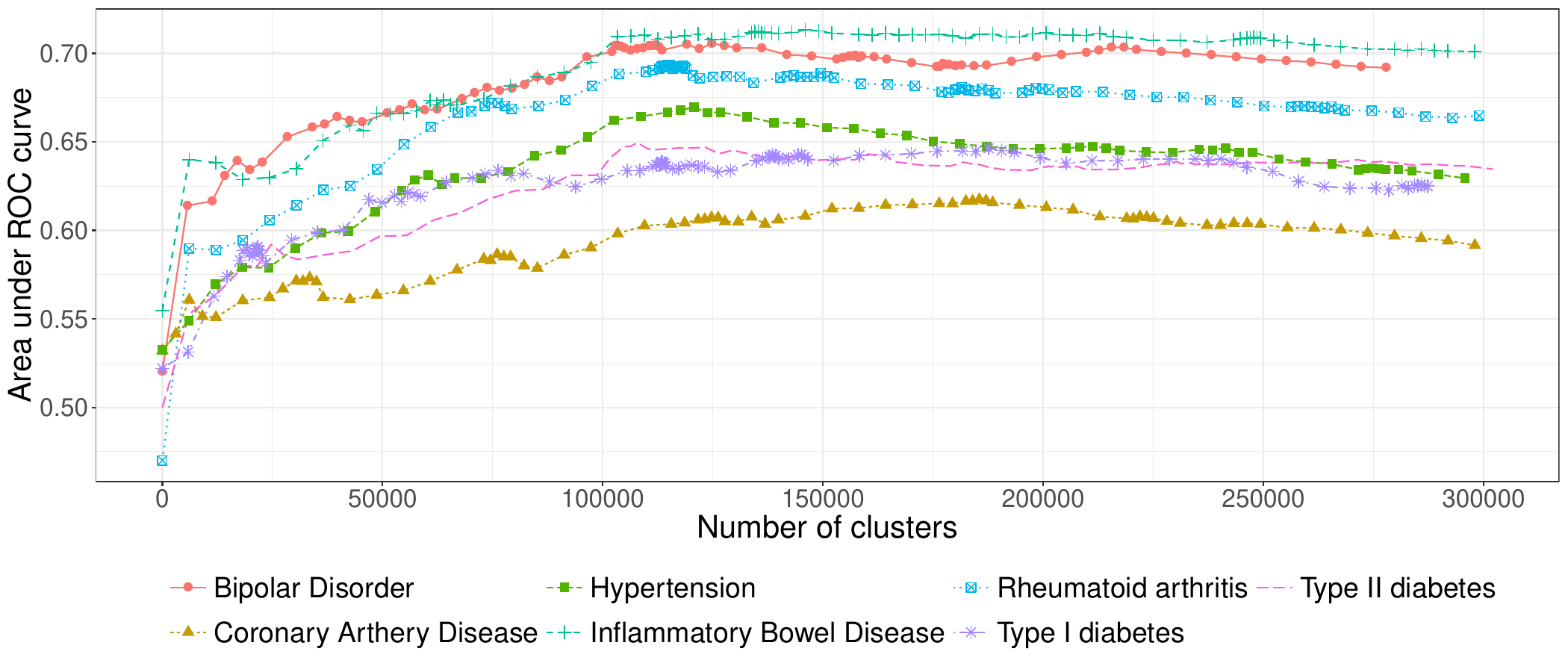}}
    \caption{AUC-ROC for different cut levels in a HAC-tree of the
      spondylitis arthritis disease (Immunochip genechip). Each point
      corresponds to an AUC value computed on a test set from a
      logistic ridge regression model for a given level in the
      constrained-HAC tree.}
    \label{fig:spondy}
\end{figure}

In contrast, the AUC-ROC from the AS data (Figure \ref{fig:spondy})
behaves differently from the WTCCC data. Predictive power is
substantially improved if \emph{aggregated-SNP} predictors are used at a
fairly high level in the hierarchical tree (7478 optimal clusters
identified by the cut level algorithm). It is relevant to note that
the pattern we observe on this real dataset is similar to the pattern
we observed in the numerical simulations, especially under the
\emph{clusSNP} scenario.

\begin{figure}[H]
  \centering
  \fbox{\includegraphics[width=.8\textwidth]{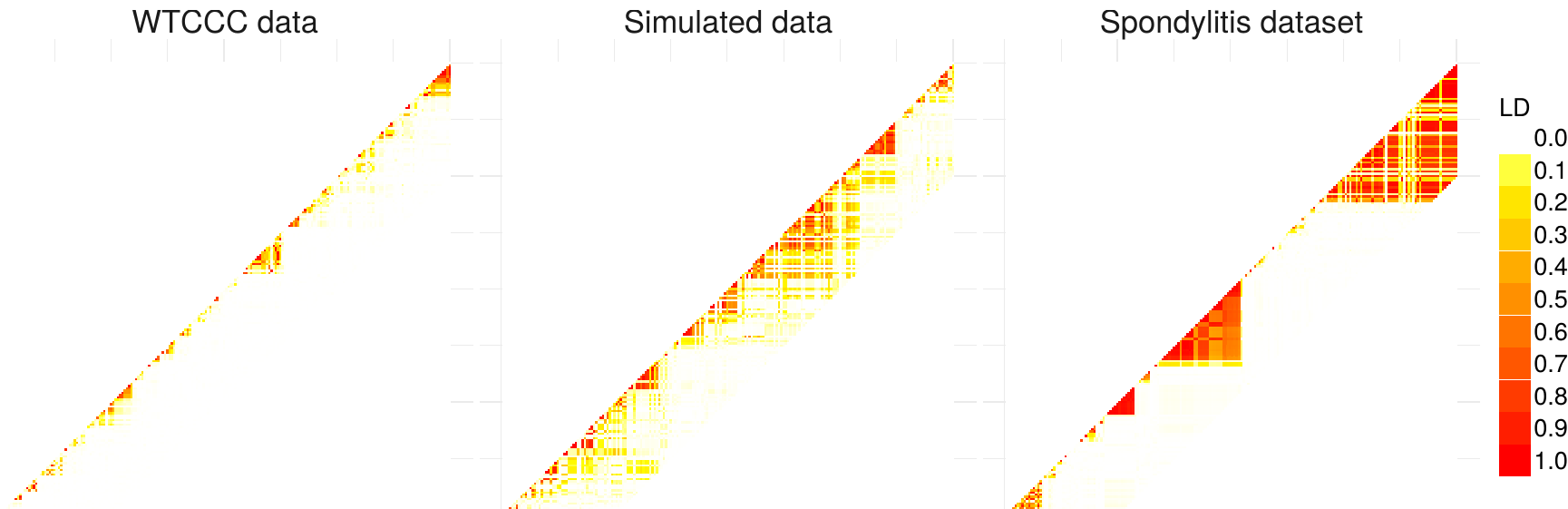}}
  \caption{Comparison of linkage disequilibrium level among SNPs for 3
    different types of dataset: WTCCC, simulated and ankylosing
    spondylitis datasets. LD computation is based on $R^2$ between SNPs.}
  \label{fig:ggheat}
\end{figure}

As we remarked concerning the WTCCC results, the algorithm identifies
a relatively high number of clusters in relation to AS and simulated
data. This difference is certainly due to the LD level among the
genetic markers in the Affymetrix GeneChip. The correlation levels
among SNPs for a given bandwith are similar between the simulated and
the AS data, but greater than for the WTCCC data
(Table~\ref{tab:marker_density} and Figure \ref{fig:ggheat}). This
suggests that there is a stronger LD pattern between blocks of SNPs in
AS and simulated data, implying that the optimal number of clusters
identified by the algorithm is dependent on the LD level among
variables.

\begin{table}[H]
\centering
        \begin{tabular}{@{} *5l @{}} 
          \toprule 
          \emph{Dataset} & \emph{SNP/kb} & Median & Mean \\   
          \midrule Simulated data & $\numprint{1.3e-27}$ &
                                                           $\numprint{1e-2}$ & 0.11 \\ 
          WTCCC data & $\numprint{7e-32}$ & $\numprint{9e-4}$ & 0.03\\ 
          AS data & $\numprint{9e-9}$ & $\numprint{3e-2}$ & 0.27\\ 
          \hline
        \end{tabular}
        \vspace{0.6cm}
        \caption{Comparison of marker density and averaged LD level between
          markers in a region of 300 SNPs for the different datasets}
        \label{tab:marker_density}
 \end{table}

\subsubsection*{GWAS analysis on AS and WTCCC datasets}

\begin{figure}[H] 
  \centering
  \fbox{\includegraphics[width=.8\textwidth]{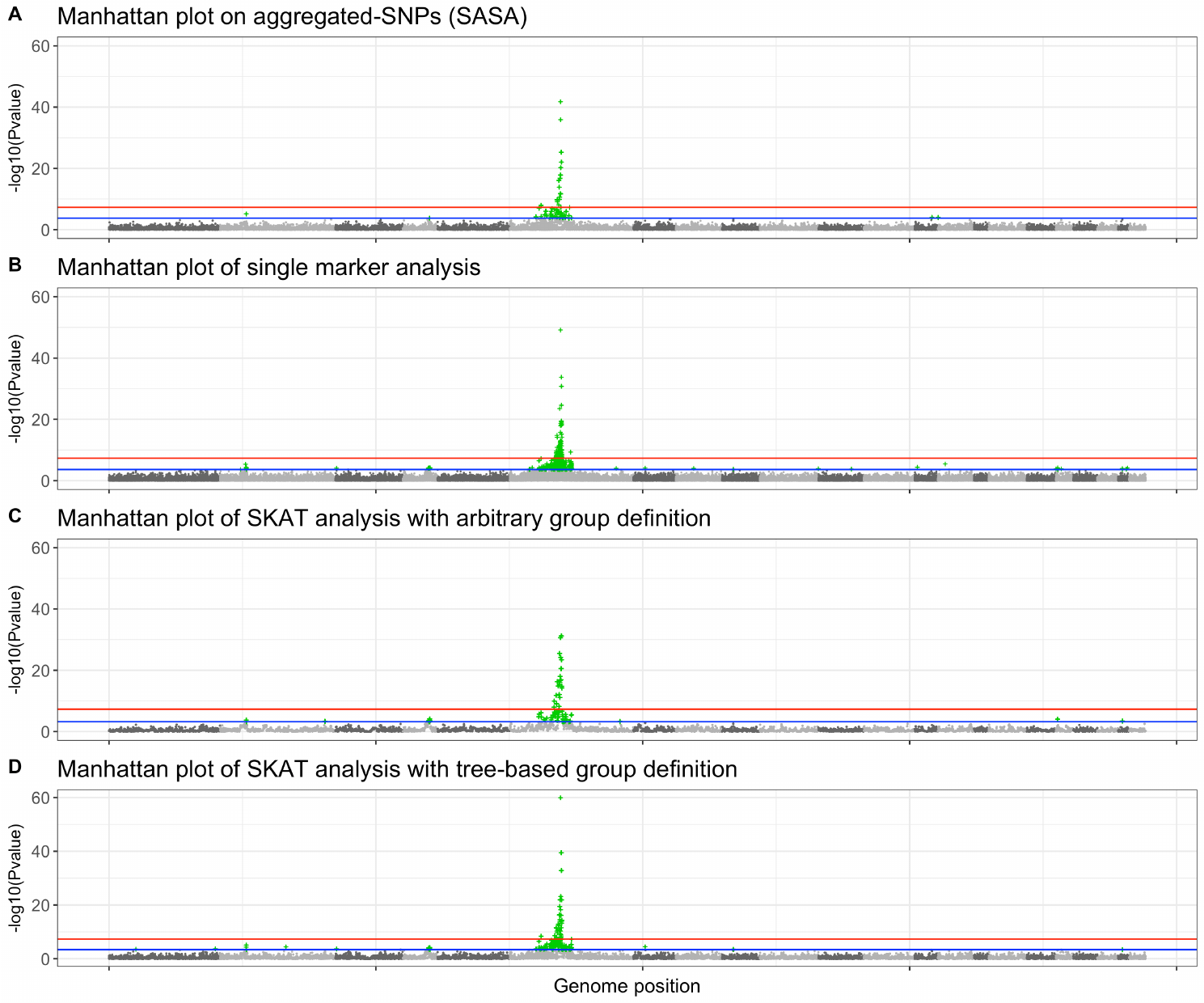}}
  \caption{Manhattan plots showing results of GWAS analysis on
    ankylosing spondylitis data. For each Manhattan plot, the
    Benjamini-Hochberg (BH) threshold is represented by the blue line and
    the Bonferroni threshold by the red line. According to the
    BH threshold, there are: (A) 64 significantly associated
    aggregated-SNPs; (B) 602 significantly associated single
    SNP; (C) 80 significantly associated
      groups of SNPs and (D) 138 significantly associated groups of SNPs.}
  \label{fig:manhattan}
\end{figure}

To evaluate the ability of our procedure to discover new associations
between SNPs and ankylosing spondylitis, we compare our procedure with
the univariate approach (SMA) and SKAT model with our group definition
and arbitrary group definition (20 SNPs). For SASA, we perform
multiple hypothesis testing on the aggregated-SNP predictors in order
to unravel significant associations with the
phenotype. Figure~\ref{fig:manhattan} presents the results of the
association analysis. For each method the logarithm of the $p$-value
of the different predictors is plotted along their position on the
genome (this plot is also known as Manhattan plot).

Either methods highlight a region on chromosome 6
strongly associated with the phenotype. This region corresponds to the
Major Histocompatibility Complex (MHC), and Human Leukocyte Antigen
(HLA) class I molecules HLA B27 belonging to this region have been
identified as a genetic risk factor associated with ankylosing
spondylitis \citep{woodrow_HLA_1978}. Our method SASA succeeds in
detecting this risk locus with a good precision, 64 aggregated-SNPs
variables are significantly associated with the phenotype compared to
602 significantly associated SNPs with the standard SMA approach.

\begin{figure}[H] 
  \centering
  \fbox{\includegraphics[width=0.9\textwidth]{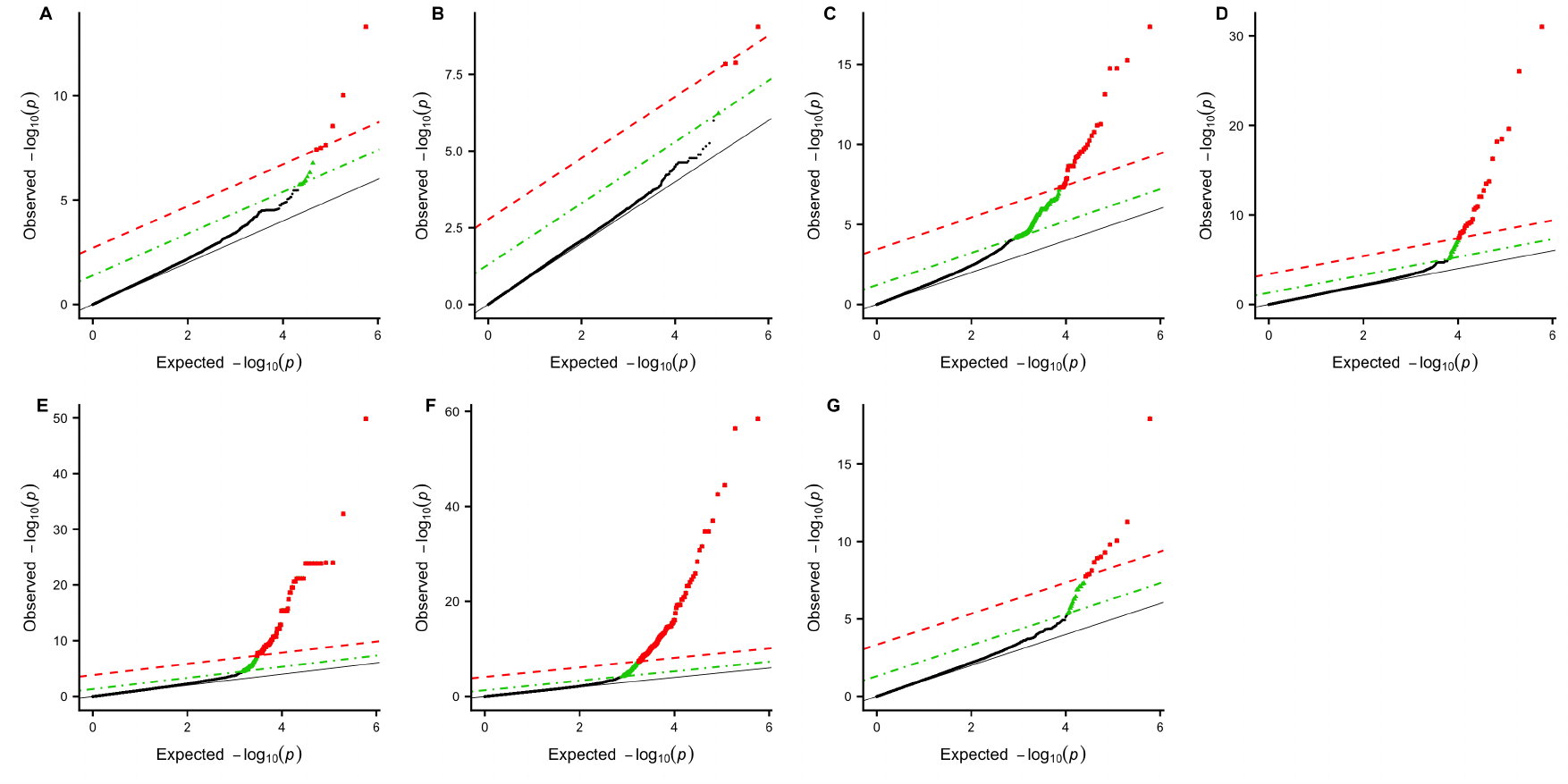}}
  \caption{Q-Q plots of group-based genome-wide analysis on WTCCC data
    using the SASA approach.  For each Manhattan plot, the
    Benjamini-Hochberg (BH) threshold is represented by the green dotted line and
    the Bonferroni threshold by the red dashed line. (A) Bipolar disorder - 13 significant
    clusters of SNPs; (B) Coronary arthery disease - 4 significant
    clusters of SNPs; (C) Inflammatory bowel disease - 356 significant
    clusters of SNPs ; (D) Hypertension - 47 significant clusters of
    SNPs ; (E) Rheumatoid arthritis - 202 significant clusters of SNPs
    ; (F) Type I diabetes - 358 significant clusters of SNPs ; (G) Type
    II diabetes - 28 significant clusters of SNPs.}
  \label{fig:qqplot}
\end{figure}

For the analysis of the WTCCC datasets, we represent the results, in
Figure~\ref{fig:qqplot}, by plotting the expected $p$-value against
the observed $p$-value (this a plot is known as Quantile-Quantile
plot). We perform the analysis using our approach SASA only.

\section*{Discussion}
\label{sec:discussions}
Overall,accounting for the linkage disequilibrium structure of
the genome and aggregating highly-correlated SNPs is seen to be a
powerful alternative to standard marker analysis in the context of
GWAS. In terms of risk prediction, our algorithm proves to be very
effective at classifying individuals given their genotype, while in
terms of the identification of loci, it shows its ability to
identify genomic regions associated with a disease with a higher
precision than standard methods.

Is is also worth mentioning that our algorithm can also accomodate
imputed variables as imputation in GWAS uses the Linkage
Disequilibrium between variables to improve the coverage of variants.
Our method being based on LD to define groups of common variants, we
expect the group structure not to be impacted by imputation.

In this work we propose a four-step method explicitly designed to
utilize the linkage disequilibrium in GWAS data. Our method combines,
on the one hand, unsupervised learning methods that cluster
correlated-SNPs, and on the other hand, supervised learning techniques
that identify the optimal number of clusters and reduce the dimension
of the predictor matrix.  We evaluated the method on numerical
simulations and real datasets and compared the results with standard
single-marker analysis and group-based approaches (SKAT\emph{tree} and
SKAT\emph{notree}). We remarked that the combination of our
aggregating function with a ridge regression model leads to a major
improvement in terms of predictive power when the linkage
disequilibrium structure is strong enough, hence suggesting the
existence of multivariate effects due to the combination of several
SNPs. These results remained consistent across a wide range of real
datasets (WTCCC and ankylosing spondylitis datasets).

In terms of the identification of associated loci in different
simulation scenarios, our method demonstrates its ability to retrieve
true causal SNPs and/or clusters of SNPs with substantially higher
precision coupled with a good power. On real GWAS data, our method has
been able to recover a genomic region associated with ankylosing
spondylitis (HLA region on chromosome 6) with a higher precision than
standard single-marker analysis.

To improve our method further, while taking into account structured input
variables in GWAS, there are different avenues that may be
explored. One avenue would involve highlighting potential non-linear
relationships between aggregated-SNPs and a response phenotype. This
could be done by making use of the continuous nature of
aggregated-SNPs variables (in contrast to the ordinal nature of
single SNP variables), by using generalized additive models
\citep{breiman_fitting_1993}, and by performing non-linear regression
using natural polynomial splines. In addition, whereas we evaluated
our method for binary traits (case-control phenotype), a possible
extension might include quantitative non-binary traits (i.e., using a
ridge regression model instead of logistic ridge regression).

\section*{Acknowledgement}

This study makes use of data generated by the Wellcome Trust
Case-Control Consortium. A full list of the investigators who
contributed to the generation of the data is available from
www.wtccc.org.uk. Funding for the project was provided by the Wellcome
Trust under award 076113, 085475 and 090355.

\section*{Availability of data and material}
A webtool to run the proposed approach have been developed and an access is available on demand. If you wish to use it, please send an email to the corresponding author. 

\section*{Ethics approval and consent to participate}
The dataset regarding ankylosing spondylitis consists of the French
subset of the large study of the International Genetics of Ankylosing
Spondylitis (IGAS) study \cite{IGAS_2013_identification}. For this subset, unrelated cases were
recruited through the Rheumatology clinic of Ambroise Par\'e Hospital
(Boulogne-Billancourt, France) or through the national self-help
patients' association: "Association Fran\c{c}aise des
Spondylarthritiques". Population-matched unrelated controls were
obtained from the "Centre d'Etude du Polymorphisme Humain", or were
recruited as healthy spouses of cases. The protocol was reviewed and
approved by the Ethics committee of the Ambroise Par\'e hospital. All
participants gave their informed consent to the study.

\bibliography{references}

\end{document}